\documentclass[11pt]{article} 

\usepackage[utf8]{inputenc} 

\usepackage{geometry} 
\usepackage{graphicx} 

\usepackage{booktabs} 
\usepackage{array} 
\usepackage{paralist} 
\usepackage{verbatim} 
\usepackage{subfig} 

\usepackage{fancyhdr} 
\usepackage{sectsty}
\allsectionsfont{\sffamily\mdseries\upshape} 

\usepackage[nottoc,notlof,notlot]{tocbibind} 
\usepackage[titles,subfigure]{tocloft} 

\usepackage{upgreek}
\usepackage{amsfonts,amssymb,amsmath}
\usepackage{mathrsfs}
\usepackage{authblk}
\usepackage[numbers,sort&compress]{natbib}

\usepackage{color}

\title{Analogue Transformations in Physics and their Application to Acoustics}
\author[1]{C. Garc\'{i}a-Meca}
\author[2]{S. Carloni}
\author[3]{C. Barcel\'{o}}
\author[4]{G. Jannes}
\author[5]{J. S\'{a}nchez-Dehesa}
\author[1]{A. Mart\'{i}nez}
\affil[1]{Nanophotonics Technology Center, Universitat Polit\`{e}cnica de Val\`{e}ncia, 46022
Valencia, Spain.}
\affil[2]{ESA – Advanced Concepts Team, ESTEC, Keplerlaan 1, Postbus 299, 2200 AG Noordwijk, The Netherlands.}
\affil[3]{Instituto de Astrof\'{i}sica de Andaluc\'{i}a (CSIC), Glorieta de la Astronom\'{i}a, 18008 Granada, Spain.}
\affil[4]{O.V. Lounasmaa Laboratory, Aalto University, 00076 Aalto, Finland.}
\affil[5]{Wave Phenomena Group, Universitat Polit\`{e}cnica de Val\`{e}ncia, 46022 Valencia, Spain.}
\date{} 

\def\x{{\mathbf x}}%
\def\v{{\mathbf v}}%
\def\bfnabla{{\mathbf \nabla}}%
\begin{document}
\maketitle

\newpage

\section*{Abstract}

Transformation optics has shaped up a revolutionary electromagnetic design paradigm,
enabling scientists to build astonishing devices such as invisibility cloaks. Unfortunately,
the application of transformation techniques to other branches of physics is often constrained
by the structure of the field equations. We develop here a complete transformation method
using the idea of analogue spacetimes. The method is general and could be considered as
a new paradigm for controlling waves in different branches of physics, from acoustics in
quantum fluids to graphene electronics. As an application, we derive an "analogue
transformation acoustics" formalism that naturally allows the use of transformations mixing
space and time or involving moving fluids, both of which were impossible with the standard
approach. To demonstrate the power of our method, we give an explicit design of a spacetime
compressor for acoustic waves and a carpet cloak for a moving aircraft.

\section*{Introduction}
Together with metamaterial science, transformation optics has dramatically improved
our control over the manipulation of electromagnetic waves. This technique provides
a way to know the properties that a medium should have in order to curve light propagation
in almost any desired way. As a consequence, it has allowed for the creation of optical
devices that were unthinkable only a decade ago~\cite{PEN06-SCI,LEO06-SCI,SCH06-SCI,SHA08-SCI,GRE09-BAMS,GEN09-NP,CHE10-NM,LEO,PEN12-SCI}.

Transformation optics is standardly described as relying on the form-invariance of Maxwell's equations~\cite{PEN06-SCI,SCH06-SCI,SHA08-SCI,GRE09-BAMS,GEN09-NP,CHE10-NM}, i.e., the fact that they have the same structure in any coordinate system~\cite{POST}.  In particular, the electromagnetic equations of a (virtual) medium $M_{\rm V}$ using a distorted (non-Cartesian) set of coordinates $S_{\rm D}$ are formally equal to those of a different (real) medium $M_{\rm R}$ in Cartesian coordinates $S_{\rm C}$. For instance, if the virtual medium is homogeneous and isotropic, it will be non-diffractive. Then, the equivalence $\{M_{\rm V},S_{\rm D}\} \sim \{M_{\rm R}, S_{\rm C}\}$ allows us to design real media that are highly non-trivial (in general, inhomogeneous and anisotropic), but nevertheless retain the non-diffractive property. 

Inspired by the success of transformation optics, scientists have tried to apply a similar procedure in other branches of physics, such as acoustics~\cite{CUM07-NJP,CHE07-APL,NOR08-JASA,CHE10-JPDAP} or quantum mechanics~\cite{ZHA08-PRL}. In these attempts, form-invariance seems to be the {\em conditio sine qua non}: without this property, a coordinate transformation cannot be reinterpreted as a certain medium. In many cases, the equations under consideration are  form-invariant only under a certain subset of transformations, limiting the technique to this subset. For instance, the acoustic equations are not invariant under general transformations that mix space and time, so one cannot use transformation techniques to design devices such as time cloaks~\cite{MCC11-JOPT,FRI12-NAT} or frequency converters~\cite{CUM11-JOPT}. Here, we show that there is a way to escape this limitation. 
Let us first describe the proposal step by step in general terms.

\section*{Results}
Consider a formal class of continuous physical systems described by two sets of fields: some {\em parameter fields} (e.g. permittivity and permeability), which describe a background medium, and some {\em dynamical fields} $\Phi$ (e.g. electric and magnetic fields), describing some specific behavior on top of the background medium which we want to manipulate in our laboratory through transformation techniques. Thus, we call \emph{laboratory space} the world where these fields exist. We assume that all these systems are described using the same standard Cartesian coordinate system $S_{\rm C}=(t,\x)$. In this way, different parameter fields always correspond to different media. In addition, imagine that the $\Phi$-equations are not form-invariant under the desired transformations and therefore we cannot directly apply the traditional transformation paradigm as in optics. 

The transformation method we propose entails the 
existence of an auxiliary {\em abstract relativistic system} (we use ``relativistic'' in the General 
Relativity sense of form-invariance under {\em any spacetime} 
coordinate transformation) which is analogue to (i.e., possesses the 
same mathematical structure as) the relevant systems in laboratory 
space. There is no need for the analogue model to have any direct 
physical meaning. In fact, this procedure is the exact reverse of what 
is done in the ``analogue gravity'' programme~\cite{BAR}: there, one 
searches for laboratory analogues of relativistic phenomena, e.g. in the 
quest to simulate Hawking radiation; here, one searches for relativistic 
analogues of laboratory phenomena. In the relativistic analogue, the 
geometric coefficients (typically, the coefficients of the metric tensor 
$g_{\mu\nu}$) play the role of parameter fields; we will denote them 
generically by the letter $C$. One specific system in laboratory space 
is mapped to one relativistic system written in an abstract coordinate system $S_{\rm AC}=(t,\x)$
(same coordinate labels as $S_{\rm C}$).

Once we find the analogue model, the method develops as follows (see Fig.~\ref{fig:Figure_1}):

\begin{figure}[ht]
\begin{center}
  \includegraphics{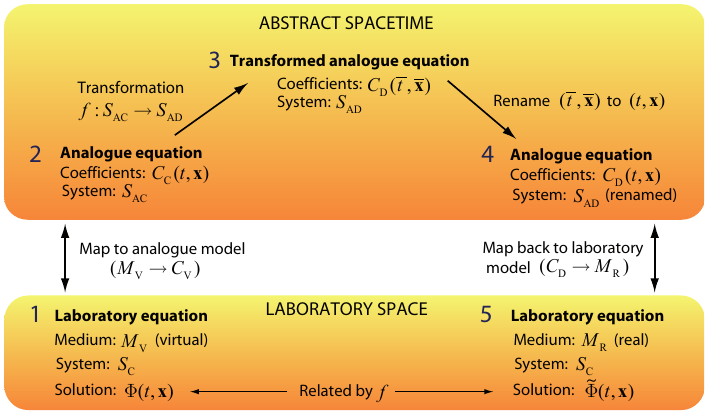}
\end{center}
  \caption{Flowchart of the proposed analogue transformation method. {In the case of acoustics, the laboratory equations correspond to equation~\eqref{potential}, which is not form-invariant, while the equations in abstract spacetime correspond to equation~\eqref{scalar-field}. The latter is form-invariant under any spacetime transformation (only the values of the metric components change depending on the coordinate system).}}
  \label{fig:Figure_1}
\end{figure}

\begin{itemize}
\item Identify a simple (and probably idealized) virtual medium $M_{\rm V}$ in laboratory space which possesses some physical or technological quality of interest (Step 1 in Fig.~\ref{fig:Figure_1}).

\item Map the laboratory equation associated with $M_{\rm V}$ to its analogue equation in the abstract spacetime written in coordinates $S_{\rm AC}$ (Step 2). The geometric coefficients $C_{\rm C}(t,\x)$ in these coordinates will be functions of the parameter fields of $M_{\rm V}$.

\item Express the analogue equation in a transformed set of coordinates $S_{\rm AD}=(\bar{t},\bar{\x})$ using the transformation $f:S_{\rm AC} \rightarrow S_{\rm AD}$ that encodes the desired distortion. {\em One can apply any transformation one likes} since the relativistic equations always maintain their form. This yields a new functional form for the geometric coefficients $C_{\rm D}(\bar{t},\bar{\x})$. Then, rename the new coordinates $(\bar{t},\bar{\x})$ to $(t,\x)$. 

\item Map the renamed equation back to laboratory space and obtain the new medium $M_{\rm R}$ associated with $C_{\rm D}(t,\x)$ (Step 5). In $M_{\rm R}$, the change induced by $f$ is ``real'', i.e., the solutions in medium $M_{\rm R}$ are related to those in medium $M_{\rm V}$ by this transformation $f$.
\end{itemize}

The reason why medium $M_R$ induces the desired change is that the solutions of the renamed equation (Step 4)
are related to those of the original analogue equation (Step 2) by $f$. Since the solutions of the equations in Steps 1 and 2
and Steps 4 and 5 are identical, this correspondence is mapped to laboratory space, i.e., the solution $\tilde{\Phi}$ associated to medium $M_{\rm R}$ is
the desired distorted version of the solution $\Phi$ associated with medium $M_{\rm V}$. 
The advantage of the analogue method proposed here is that the transformation approaches can now be based on any symmetry that leaves the form of the \emph{analogue equations} invariant, rather than the restricted set which corresponds to the \emph{original equations} in laboratory space.

As in any transformational approach, the medium $M_{\rm R}$ must be flexible enough to
endow the laboratory equation with sufficient degrees of freedom (through flexible parameter
fields) to reproduce the values of $C_{\rm D}$ associated with the transformations of interest. This will
typically require that the system includes moving media and metamaterials.

Let us now apply our new method to acoustics. 
The standard approach to transformation acoustics (STA) is based on the wave equation
for the pressure perturbations $p$ of a fluid medium

\begin{align}
\label{pressure}
\ddot p= B \partial_i \rho^{ij} \partial_j p~.
\end{align}

Here, $B$ is the bulk modulus and $\rho^{ij}$ the anisotropic inverse matrix density of the background fluid. We use latin spatial indices ($i,j$) and Greek spacetime indices ($\mu, \nu$, with $x^0=t$). $B$ and $\rho^{ij}$ are the parameter fields, while $p$ is the dynamical field.
From the transformational physics point of view, equation~\eqref{pressure} has a set of drawbacks.
i) Clearly, its form is non-invariant under coordinate transformations that mix space and time. ii) It
cannot deal with fluids flowing with a non-zero background velocity, which however is the case in many interesting physical
situations. iii) To derive equation~\eqref{pressure}, a necessary assumption is that the background (static) pressure is
homogeneous~\cite{BER46-JASA}. Thus, it cannot be used for fluids with significant pressure gradients,
such as when gravitational forces are relevant. iv) Three-dimensional transformations that require a local increase
of the speed of sound in $M_{\rm R}$, also require an increase of its density. This is difficult to achieve with current
metamaterials in airborne sound~\cite{TOR06-PRL,TOR06-PRB}.  

The proposed method aims in the first place at overcoming the first drawback. 
Rather than starting from equation~\eqref{pressure}, one can look at the field of analogue gravity, 
where the formulation of a form-invariant equation has become a standard procedure.
In terms of the velocity potential $\phi$ (the velocity perturbation is then $\v_p=\bfnabla \phi$), the equation for the acoustic perturbations of a barotropic and irrotational fluid is written as~\cite{Unruh,Visser,BAR}
\begin{align}
\label{potential}
-\partial_t\left(\rho{c}^{-2}\left(\partial_t\phi +\mathbf{v}\cdot\nabla\phi\right)\right)+\nabla\cdot\left(\rho\nabla\phi-\rho c^{-2}\left(\partial_t\phi + \mathbf{v}\cdot\nabla\phi\right)\mathbf{v}\right)=0,
\end{align}
with $\rho$ the mass density, $c$ the speed of sound, and $\mathbf{v}$ the fluid background velocity.
This apparently innocent reformulation will already turn out to have several important side-benefits. However, so far, this equation is still not form-invariant under transformations that mix space
and time. Therefore, we now apply one of the crucial insights of analogue gravity: 
equation~\eqref{potential} is structurally identical to the equation describing a relativistic massless scalar field
over a curved spacetime~\cite{BAR}
\begin{align}
\label{scalar-field}
{1\over \sqrt{-g} } \partial_\mu \sqrt{-g} g^{\mu\nu} \partial_\nu \phi=0~ ,
\end{align}
with the (acoustic) metric
\begin{eqnarray}
\label{acoustic-metric}
g_{\mu\nu} = {\rho \over c}
\left(
\begin{array}{ccc}
-(c^2- v^2) & \vdots & -v^i \\
.......... & . & ...... \\
-v^i & \vdots & \delta_{ij}
\end{array}
\right)~,
\end{eqnarray}
and $g$ the metric determinant. {Note that in this definition the choice of units can be considered irrelevant (see Supplementary Information).} Equation~\eqref{scalar-field} retains its form upon any coordinate
transformation. Thus, using equation~\eqref{potential} as the laboratory equation and equation~\eqref{scalar-field}
as the analogue equation, we can apply the proposed procedure. The virtual medium $M_{\rm V}$ in step 1 of Fig.~\ref{fig:Figure_1}
will be characterized by some parameters $\rho_{\rm V}$, $c_{\rm V}$ and $\mathbf{v}_{\rm V}$, and the real medium $M_{\rm R}$ in step 5
by other parameters $\rho_{\rm R}$, $c_{\rm R}$ and $\mathbf{v}_{\rm R}$. According to equation~\eqref{scalar-field}, the coefficients
$C_{\rm V}$ associated to $M_{\rm V}$ are $\sqrt{-g}g^{\mu\nu}$, with equation~\eqref{acoustic-metric} particularized for
$\rho_{\rm V}$, $c_{\rm V}$ and $\mathbf{v}_{\rm V}$. After applying all steps in Fig.~\ref{fig:Figure_1}, one obtains the relation
between the parameter fields of $M_{\rm V}$ and the new medium $M_{\rm R}$ that reproduces the effect of the coordinate
change.

\begin{figure}[ht]
\begin{center}
  \includegraphics{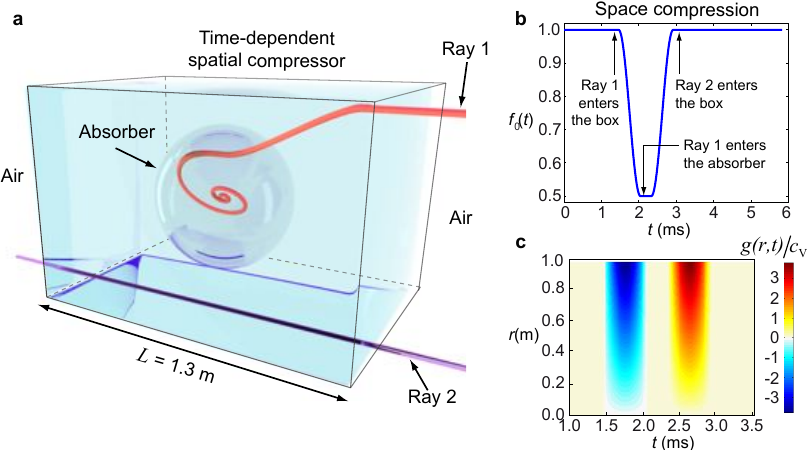}
\end{center}
  \caption{Time-dependent spatial compressor and omnidirectional absorber. {The absorber has a spherical shape with radius $R=0.29$ m, and is characterized by a static refractive index (relative to the background medium surrounding the compressor) given by $n=0.5 R/r$.} The trajectories of two acoustic rays are simulated.
After Ray 1 enters the box, the compression starts, according to the compression function $f_0(t)$ (depicted on the left). Ray 1 approaches the absorber, which traps it. Ray 2 feels no compression since $f_0(t)=1$ during the time it goes through the box. Thus, Ray 2 follows a straight line.  See Supplementary Information for more details.
{(a) Simulated acoustic rays. (b) Compression function $f_0(t)$. (c) Function $g(r,t)$.}}
  \label{fig:Figure_2}
\end{figure}

As a result, many interesting transformations that involve mixing space and time, which could not be addressed in STA, now become possible (for instance, all those that mix time with one spatial variable -- see Supplementary Information). A concrete example consists of a time-dependent compression of space which acts only inside a three-dimensional box (the compressor). Such a compressor could be used, for instance, to select which rays are absorbed by a static omnidirectional absorber placed inside the box. {The simulated performance of a specific configuration of the compressor-absorber device is shown in Fig.~\ref{fig:Figure_2} (see the Methods section and the Supplementary Information for simulation details). It is based on the transformation $\bar{t}=t$ and $\bar{x}^i=x^if_0(t)$. According to the proposed method, this transformation can be implemented by a set of parameters given by $c_{\rm R}/c_{\rm V}=\rho_{\rm V}/\rho_{\rm R}=f_0(t)$ and $\mathbf{v}_{\rm R}=g(r,t)/f_0(t)\hat{r}$, with $g(r,t)=r\partial_t f_0(t)$ and $\mathbf{r}=r\hat{r}$ being the position vector. Both $f_0(t)$ and $g(r,t)$ (normalized to $c_{\rm V}$) are depicted in Fig.~\ref{fig:Figure_2}.}

\begin{figure}[ht]
\begin{center}
  \includegraphics{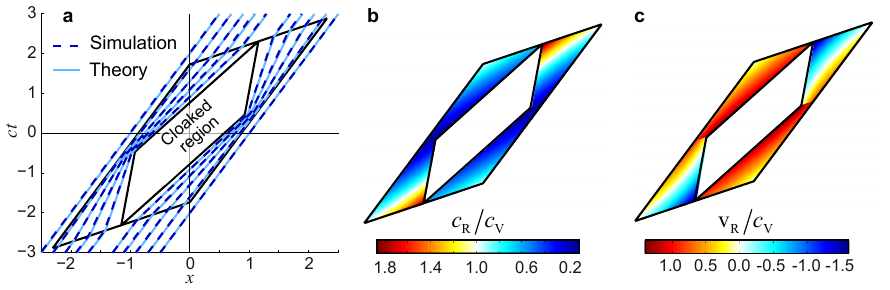}
\end{center}
  \caption{{Acoustic spacetime cloak. (a) Simulated acoustic rays. As can be seen, sound propagation is speeded up or slowed down in order to hide any acoustic event belonging to the cloaked region. Rays exit the cloak as if they had just passed through the background fluid. (b) Sound speed and (c) background velocity $\mathbf{v}_{\rm R}={\rm v}_{\rm R}\hat{x}$ of the cloak.}}
  \label{fig:Figure_3}
\end{figure}

{Another example is the acoustic counterpart of the spacetime cloak reported recently for electromagnetic waves~\cite{MCC11-JOPT}. Unlike static invisibility cloaks, this device conceals only a certain set of spacetime events occurring during a limited time interval. To show how the proposed method allows for designing an acoustic spacetime cloak, we consider the transformation given by equations~(20)-(23) in Ref.~\cite{MCC11-JOPT} (the transformation can also be found in the Supplementary Information). This transformation mixes time with one spatial variable. Therefore, the material parameters associated with the cloak can be obtained by substituting the mentioned transformation into equations~(58)-(60) of the Supplementary Information. The resulting parameters are shown in Fig.~\ref{fig:Figure_3}. A set of acoustic rays propagating through such a medium were simulated and compared to the expected trajectories, finding and excellent agreement (see Fig.~\ref{fig:Figure_3}).}

\begin{figure}[ht]
\begin{center}
  \includegraphics{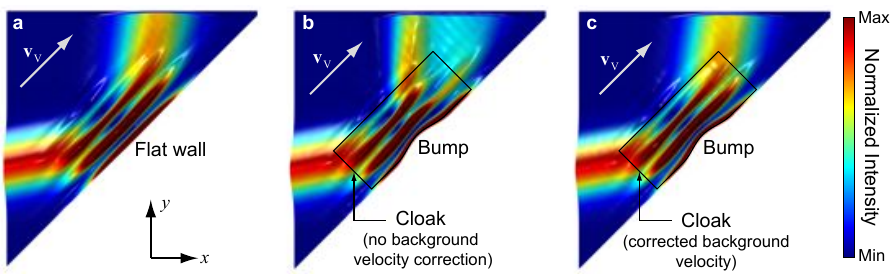}
\end{center}
  \caption{Wave intensity around a carpet cloak for acoustic waves in moving fluids. A Gaussian beam initially propagates along the positive x-direction.
The wave is immersed in a fluid moving parallel to the aircraft wall (with speed $0.35c_{\rm V}$), and thus suffers a dragging effect that deviates it from its original trajectory.
(a) Wave impinging onto a flat wall. (b) Wave impinges onto a wall with a bump surrounded by a carpet cloak
within which the background fluid flows with the same velocity as outside the cloak. The wave outside
the cloak is significantly distorted compared to the flat-wall case. (c) As in (b), but now the background fluid flows with a modified velocity calculated with the
analogue transformation method. Outside the cloak, the wave recovers the form it had in the flat-wall case: the bump is perfectly cloaked. See Supplementary Information for more details.}
  \label{fig:Figure_4}
\end{figure}

As a side-benefit of the above procedure, we have automatically solved the other main drawbacks associated to equation~\eqref{pressure}: equations~\eqref{potential} and \eqref{scalar-field} do not require the background configuration to be of constant pressure, and can deal with non-zero velocity background flows, making them much more generally applicable. Thus, we can now for example perform transformation acoustics with waves propagating in moving fluids, a common situation in aeronautics. Imagine, for instance, that we want to cloak a bump in an aircraft. In general, if the aircraft is moving with respect to the surrounding air, a traditional static cloaking device~\cite{LI08-PRL,POP11-PRL} will fail to cloak the bump properly, because STA does not take into account that the background fluid velocity $\mathbf{v}_{\rm R}$ must also be adapted suitably. Instead, the proposed analogue transformation method allows us to obtain the required transformation of $\mathbf{v}_{\rm R}$ (see Fig.~\ref{fig:Figure_4}).
It is worth mentioning that, although we have theoretically solved the problem, the actual realization of the cloaking device should take into account other issues like, for example, aerodynamic constraints. 
{It is worth mentioning that the designed carpet cloak does not introduce reflections, since the employed transformation is continuous at the boundaries. To further verify this fact, we calculated in COMSOL the ratio between the acoustic intensity (power) reflected to the left by the carpet cloak ($I_r$) and the intensity of the input wave ($I_i$), finding a value of $I_r/I_i \approx 10^{-4}$ (within numerical error). Note that, because one has a moving background, a Doppler effect on the acoustic wave is expected. This effect appears, however, equally in presence and absence of the carpet cloak, which does not further contribute to the evolution of the acoustic signal. The acoustic wave outside the carpet cloak is exactly the same as in the case in which it impinges onto a flat wall without carpet cloak. Thus, the frequency of the transmitted wave (i.e., reflected upwards by the wall with the bump surrounded by the cloak) is the same as the frequency of the input wave.}

In addition, STA prescribes for the case of three-dimensional transformations that sound speed and density must be increased or decreased simultaneously at each point, while the opposite behavior is obtained with the analogue approach (see Supplementary Information). This latter requirement is more in line with current state-of-the-art acoustic metafluids. These are typically built by putting a lattice of rigid objects in air, and (unlike in natural fluids) an increase of density corresponds to a decrease of the speed of sound, and vice versa~\cite{TOR06-PRL,TOR06-PRB}. 
Thus, we expect the construction of metamaterial devices designed with the new method to be easier.

{Finally, note that the equations used in the text do not contain any approximation apart the ones necessary to obtain equations~\eqref{pressure} and \eqref{potential} from basic fluid dynamics principles. Therefore, the analogue transformations method works both in the wave and eikonal regimes. On the other hand, as mentioned above, in some cases the acoustic parameters transform differently depending on whether one works with equation~\eqref{pressure} or equation~\eqref{potential}. However, both values are correct as long as the assumptions used to obtain the corresponding wave equation are valid. }

\section*{Discussion}
Given that we have described our method in abstract terms, it is natural to ask how general this approach is. The answer relies on our ability to construct an analogue theory that mirrors the properties of the relevant equations. Two concrete examples in this respect are the following. First, phonons in Bose-Einstein condensates provide an analogue model of a massless scalar field~\cite{garay,steinhauer}. Second, graphene provides an analogue model of relativistic electrons~\cite{castro}, so this method could become a powerful tool to control electronic propagation and confinement in a graphene sheet. In practice, the local manipulability of the propagating (meta)media will be crucial. In the case of graphene, the conductivity could be controlled by introducing curvature~\cite{cortijo} or by varying the chemical potential~\cite{VAK11-SCI}. More in general, the method presented here can be applied to any system which 
provides an analogue model of a relativistic field, i.e. scalar fields, electromagnetic
fields, Dirac and Weyl spinorial fields and even spin-2 (gravitational) fields \cite{BAR}. In all these examples, the problem of the non-form-invariance of the equations describing a given system can be circumvented through the construction of an analogue theory. All in all, our results indicate that the idea underneath transformation optics has an even far richer scope than initially foreseen. 

\section*{Methods}
Numerical simulations have been performed with the commercially available COMSOL Multiphysics simulation software,
which is based on the finite element method. For full-wave simulations, the velocity potential wave equation has been
numerically solved using COMSOL's acoustic module.

The trajectories followed by acoustic rays in the geometrical approximation have been calculated by solving Hamilton's
equations:
\begin{align}
\frac{d\mathbf{k}}{dt}&=-\frac{\partial{H}}{\partial\mathbf{q}} \\
\frac{d\mathbf{q}}{dt}&=\frac{\partial{H}}{\partial\mathbf{k}}
\end{align}
In our case, the Hamiltonian equals the angular frequency, which can be obtained from the dispersion relation
associated to the velocity potential wave equation under the assumption of plane-wave solutions of the form
$\phi=\phi_0e^{j(\mathbf{k}\cdot\mathbf{r}-\omega{t})}$. Thus, it can be shown that the sought Hamiltonian is ($k=|\mathbf{k}|$)
\begin{align}
H=ck+\mathbf{v}\cdot\mathbf{k}.
\end{align}
%
\section*{Supplementary Material}
The supplementary information is organised as follows. First, we will describe in some more detail the application of the analogue transformation method to the case of acoustics. Next, we will apply this to spatial transformations and compare the results of the analogue transformation method with those obtained from the standard formalism for transformation acoustics. In the third section, we will demonstrate that, contrarily to the standard scheme, analogue transformation acoustics can deal with spacetime transformations that mix space and time. Finally, in the fourth section, we will give some numerical examples that illustrate our findings.
%
\section{Analogue transformation acoustics (ATA)}
\begin{figure}
\begin{center}
  \includegraphics{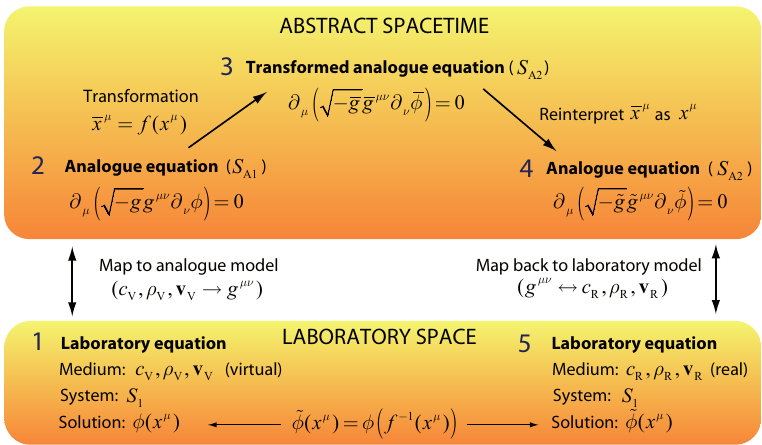}
\end{center}
  \caption{Analogue Transformation Acoustic (ATA) scheme. As in the main text, $\mu, \nu$ represent spacetime indices ($\mu, \nu=0-3$, with $x^0=t$). $\partial_\mu$ is the partial derivative with respect to $x^\mu$, and the Einstein summation convention over repeated indices is employed.}
  \label{fig:Proposal}
\end{figure}

Let us describe more extensively the details of Analogue Transformation Acoustics. In the definition of ATA
we will refer to Fig.\ref{fig:Proposal}, which specializes the general analogue transformation method to the case of acoustics.

We start from the definition of the analogue model for acoustics. As mentioned in the main text, this is a very well
studied field and full details can be found in \cite{Unruh,Visser,BAR}. The equation for the velocity potential
$\phi$ in a barotropic, inviscid and irrotational fluid with density $\rho$, sound speed $c$, and background
velocity $\mathbf{v}$ is:
\begin{align}
\label{wave_moving}
-\partial_t\left(\rho{c}^{-2}\left(\partial_t\phi +\mathbf{v}\cdot\nabla\phi\right)\right)+\nabla\cdot\left(\rho\nabla\phi-\rho c^{-2}\left(\partial_t\phi + \mathbf{v}\cdot\nabla\phi\right)\mathbf{v}\right)=0.
\end{align}
It is a known result from analogue gravity that this equation can be written as the equation of motion
of a scalar field $\phi$ propagating in a (3+1)-dimensional pseudo--Riemannian manifold (Equation 2
in Fig. \ref{fig:Proposal}), also called d'Alembert, Laplace-Beltrami or (massless) Klein-Gordon equation:
\begin{equation} \label{dAlembertian}
{1\over\sqrt{-g}} \partial_\mu \left( \sqrt{-g} \; g^{\mu\nu} \; \partial_\nu \phi \right)=0,
\end{equation}
provided that the contravariant metric has components~\cite{BAR} 
\begin{equation}\label{Ametric}
g^{\mu\nu}(t,\x) \equiv 
{1\over \rho c}
\left( \begin{matrix}-1&\vdots&-v^j\cr
               \cdots\cdots&\cdot&\cdots\cdots\cdots\cdots\cr
	       -v^i&\vdots&(c^2 \; \delta^{ij} - v^i \; v^j )\cr \end{matrix} 
\right),	       
\end{equation}
or, lowering the indices to obtain the (covariant) metric itself
\begin{equation}\label{metric-bis}
g_{\mu\nu}(t,\x) \equiv 
{\rho \over  c}
\left( \begin{matrix}-(c^2-v^2)&\vdots&-v^j\cr
               \cdots\cdots\cdots\cdots&\cdot&\cdots\cdots\cr
	       -v^i&\vdots&\delta_{ij}\cr \end{matrix}
\right),	       
\end{equation}
with
\begin{equation}
\sqrt{-g} = {\rho^2\over c}.
\end{equation}
It is important to stress here that $g_{\mu\nu}$ does not represent the metric of the laboratory spacetime
but corresponds to the abstract spacetime in which the analogue
model is defined. 

{Note that since the acoustic equation can always be multiplied by a constant term $\rho_{\rm ref}/c_{\rm ref}$, the acoustic metric can be defined with any dimension. In fact, using coordinates $x^0=c_{\rm ref}t,x,y,z$, the Klein-Gordon equation remains the same, while 
\begin{eqnarray}
g_{\mu\nu} = {\rho \over c}{c_{\rm ref} \over \rho_{\rm ref}}
\left(
\begin{array}{ccc}
-(c^2- v^2)/c^2_{\rm ref} & \vdots & -v^i/c_{\rm ref} \\
.......... & . & ...... \\
-v^i/c_{\rm ref} & \vdots & \delta_{ij}
\end{array}
\right)
\end{eqnarray}
i.e., a non-dimensional metric. Regarding the calculations in the paper these reference constants are irrelevant and in the following, for the sake of simplicity, we will not include them.}

The analogy described by the previous equations is only valid in Cartesian coordinates.
However, it is sometimes useful to work in other coordinate systems.
In a general coordinate system,  equation~\eqref{wave_moving} will not have the form of 
equation~\eqref{dAlembertian} (otherwise we would not need the analogy). Nonetheless, introducing
the three-dimensional spatial metric $\gamma_{ij}$ ($\gamma^{ij}$ being its inverse and $\gamma$
its determinant) and using general spatial coordinates (fixing time to be
laboratory time), it can be shown that, if $\gamma$ is time-independent, equation~\eqref{wave_moving}
is still equivalent to equation~\eqref{dAlembertian} if we write the metric as
\begin{eqnarray}
g_{\mu\nu} = {\rho \over c}
\left(
\begin{array}{cccc}
-c^2+v^iv^j\gamma_{ij} &\vdots & -v^j\gamma_{ij}\\
................... & . & ........... \\
-v^j\gamma_{ij} &  \vdots & \gamma_{ij} & \\
\end{array}
\right)~,
\end{eqnarray}
\begin{eqnarray} \label{inverse_4metric}
g^{\mu\nu} = \frac{1}{\rho c}
\left(
\begin{array}{ccc}
-1 & \vdots & -v^i \\
... & . & ................ \\
-v^i & \vdots & c^2\gamma^{ij}-v^iv^j
\end{array}
\right)~,
\end{eqnarray}
\begin{eqnarray}\label{metric_determinant}
\sqrt{-g}=\sqrt{\gamma} {\rho^2 \over c}~.
\end{eqnarray}

These relations between metric and medium parameters connect the equations in laboratory space 
with those in the abstract spacetime, i.e., the equations in step 1 with those in step 2, and the equations in step 4 with those in step 5 in Fig. \ref{fig:Proposal}.
The method is composed of the following steps:
\begin{itemize}
\item  Start from a (usually simple) virtual medium characterized by parameters $\rho_{\rm V}$, $c_{\rm V}$, and $\mathbf{v}_{\rm V}$
and express the laboratory equation in a coordinate system $S_1$ for which the analogy holds.
\item  Using equation~\eqref{inverse_4metric} particularized for the parameters of the virtual medium, derive its analogue model,
which is now a covariant equation in the abstract spacetime, expressed in a coordinate system $S_{\rm A1}$.
\item  Perform a coordinate transformation $\bar{x}^{\mu}=f(x^{\mu})$ from system $S_{\rm A1}$ to another system $S_{\rm A2}$
to obtain the new metric of step 3, which follows from that in step 2 by using standard tensorial transformation rules 
\begin{equation} \label{metric_transformation}
\bar{g}^{\bar{\mu}\bar{\nu}}=\Lambda^{\bar{\mu}}_{\mu}\Lambda^{\bar{\nu}}_{\nu}g^{\mu\nu},
\end{equation}
where $\Lambda^{\bar{\mu}}_{\mu}=\partial{\bar{x}^{\bar{\mu}}}/\partial{x^{\mu}}$.
\item  Rename $\bar{x}^{\mu}$ to $x^{\mu}$ to obtain the equation in step 4.
\item  Derive the acoustic parameters $\rho_{\rm R}$, $c_{\rm R}$, and $\mathbf{v}_{\rm R}$ of the real medium that
corresponds to the equation in step 4 by using equation~\eqref{inverse_4metric}.
\end{itemize}
It is worth noting that, for the sake of simplicity, the last two points are carried out in a slightly different way.
First, we directly write equation in step 4 as a function of the parameters of $M_R$. In fact, according to equation~\eqref{inverse_4metric},
the metric $\tilde{g}^{\mu\nu}$ in the equation of step 4 should be of the form
\begin{eqnarray}
\tilde{g}^{\mu\nu} = \frac{1}{\rho_{\rm R} c_{\rm R}}
\left(
\begin{array}{ccc}
-1 & \vdots & -v_{\rm R}^i \\
... & . & ................ \\
-v_{\rm R}^i & \vdots & c_{\rm R}^2\tilde{\gamma}^{ij}-v_{\rm R}^iv_{\rm R}^j
\end{array}
\right)~,
\end{eqnarray}
Then, we force the equations of steps 3 and 4 to be equal (after relabeling $\bar{x}^{\mu}$ to $x^{\mu}$ in the expression for $\bar{g}^{\mu\nu}$), which implies that
\begin{equation} \label{ATA_condition}
\sqrt{-\bar{g}}\bar{g}^{\mu\nu}=\sqrt{-\tilde{g}}\tilde{g}^{\mu\nu}.
\end{equation}
From this equation we obtain the relation between the material parameters of the virtual and real media.

Since equation~\eqref{wave_moving} only considers isotropic fluids, the metric in equation~\eqref{metric-bis} is spatially isotropic, and we cannot implement spatially anisotropic transformations. For that, one would need to deal with anisotropic fluids and
incorporate anisotropy, e.g. through a homogenization procedure~\cite{TOR09-PRB}. In that case, it will be necessary to use the additional parameters of the anisotropic fluid on top of $\rho$, $c$ and $\mathbf{v}$. However, all transformations giving rise to an isotropic spatial part of the metric are already implementable. This includes all spatially conformal and, to a good approximation, quasi-conformal
transformations. Also, one can work out many interesting situations involving mixing space and time.
For instance, all transformations mixing time with one spatial variable are possible. Note that, as in transformation optics, in this case the implementation of spacetime transformations is related to moving media. 

In the following sections, we will study all these cases in detail to demonstrate these assertions.

\section{Spatial transformations}

As a first example of how ATA works, we will  consider purely spatial transformations i.e. transformations of the type 
$\bar{t}=t$ and $\bar{\mathbf{x}}=f(\mathbf{x})$. These transformations will allow to test the method described in the previous 
section and to compare the results with standard transformation acoustics. Since the wave equation for the velocity potential in its standard form does not account for anisotropy, we will restrict ourselves to conformal transformations, which preserve isotropy. We will
consider three different cases: two-dimensional (2D) conformal mappings and static background, three-dimensional (3D)
conformal mappings and static background, and general conformal mappings and moving background. 

Before deriving the material properties for these cases, we will first briefly review the standard method
for transformation acoustics so that we can compare the results of both approaches in subsequent sections\footnote{{Note that by STA we refer to the application of direct transformations to the pressure wave equation and by ATA we mean the combined use of the velocity potential wave equation and the analogue transformations method). Our comparison focuses only on differences of the results of the two methods.}}
\subsection{Standard Transformation Acoustics (STA)}
STA is essentially inspired by transformation optics. 
However, differently from the latter, STA is based on a second-order differential equation rather than
a set of first-order ones. This second order equation is the pressure wave equation and  can be proven to be
invariant under pure spatial transformations. The success of STA depends on this property. As mentioned above,
we will consider only isotropic media, for which the pressure wave equation reads in arbitrary spatial coordinates
\begin{align}
\label{wave_anisotropic}
\partial_t^2{p}=\frac{B_{\rm V}}{\sqrt{\gamma}}\partial_i\left(\frac{1}{\rho_{\rm V}}\sqrt{\gamma}\gamma^{ij}\partial_jp\right),
\end{align}
where $p$ is the (acoustic) first-order pressure perturbation, $B_{\rm V}$ the bulk modulus, and $\rho_{\rm V}$ the mass density.
We consider this equation to be the one representing the initial virtual space. For simplicity, we choose a simple
coordinate system $S_1$ with an associated spatial metric of the form
\begin{align}
\label{metric0}
\gamma_{ij}&=\Omega_{\rm R}\delta_{ij},
\end{align}
where $\Omega_{\rm R}$ is a function of the spatial coordinates. After a transformation the wave equation can be expressed as
\begin{align}
\label{Wave_equation_pressure_virtual}
\partial_t^2{\bar{p}}&=\frac{B_{\rm V}}{\sqrt{\bar{\gamma}}}\partial_i\left(\frac{1}{\rho_{\rm V}}\sqrt{\bar{\gamma}}\bar{\gamma}^{ij}\partial_j\bar{p}\right),
\end{align}
Restricting ourselves to conformal transformations in order to preserve isotropy, the metric $\bar{\gamma}_{ij}$
will have the form
\begin{align}
\label{metric1}
\bar{\gamma}_{ij}&=\Omega_{\rm V}\delta_{ij} \Rightarrow \bar{\gamma}=\Omega_{\rm V}^3,
\end{align}
where $\Omega_{\rm V}$ is also a function of the spatial coordinates.
Now we consider another (real) space filled with an isotropic medium characterized by parameters $\rho_{\rm R}$ and $B_{\rm R}$.
In this case the wave equation is 
\begin{align} \label{Wave_equation_pressure_real}
\partial_t^2{\tilde{p}}&=\frac{B_{\rm R}}{\sqrt{\tilde{\gamma}}}\partial_i\left(\frac{1}{\rho_{\rm R}}\sqrt{\tilde{\gamma}}\tilde{\gamma}^{ij}\partial_j\tilde{p}\right).
\end{align}
The coordinate system in real space is also $S_1$. Thus, the metric $\tilde{\gamma}_{ij}$ is
\begin{align}
\label{metric2}
\tilde{\gamma}_{ij}&=\Omega_{\rm R}\delta_{ij} \Rightarrow \tilde{\gamma}=\Omega_{\rm R}^3.
\end{align}
Equations~\eqref{Wave_equation_pressure_virtual} and \eqref{Wave_equation_pressure_real} are mathematically identical if
\begin{align}
\label{Bulk_pressure2}
B_{\rm R}&=B_{\rm V}\frac{\Omega_{\rm R}^{3/2}}{\Omega_{\rm V}^{3/2}},\\
\label{Density_pressure2}
\rho_{\rm R}&=\rho_{\rm V}\frac{\Omega_{\rm R}^{1/2}}{\Omega_{\rm V}^{1/2}}.
\end{align}
The relation between the sound speed in virtual and real spaces can be derived using the well-known expression
$c=\sqrt{B/\rho}$, yielding
\begin{align}
\label{Speed_pressure}
c_{\rm R}^2&=\frac{B_{\rm R}}{\rho_{\rm R}}=c_{\rm V}^2\frac{\Omega_{\rm R}}{\Omega_{\rm V}}.
\end{align}
As we can see, the mass density and speed of sound of virtual space are multiplied by the same factor,
i.e., they both decrease or increase simultaneously. However, as pointed out in the main text, standard
artificial composites for airborne sound follow a different behavior, with sound speed increasing when the
density decreases and vice versa~\cite{TOR06-PRL,TOR06-PRB}. This fact may complicate the implementation
of the required properties to achieve acoustic devices based on this transformation acoustics method.

For 2D transformations, the required material parameters in real space are simpler and it turns
out that they coincide with the ones derived from ATA (see next section). 2D conformal mappings
are interesting because any (analytic) function of complex variables defines a conformal mapping.
In addition, there exist several efficient numerical methods for calculating 2D quasi-conformal mappings.
To analyze this particular case, we will assume for simplicity that $S_1$ is the Cartesian system,
so that $\gamma_{ij}=\tilde{\gamma}_{ij}=\delta_{ij}$. In addition, if we leave the $z$ variable
unchanged so that
\begin{align} \label{2D_transformation}
\nonumber
\bar{x}&=\bar{x}(x,y)\\
\bar{y}&=\bar{y}(x,y)\\
\bar{z}&=z \nonumber
\end{align}
and use the properties $\partial{x}/\partial{\bar{x}}=\partial{y}/\partial{\bar{y}}$ and
$\partial{x}/\partial{\bar{y}}=-\partial{y}/\partial{\bar{x}}$ of a conformal transformation, we obtain
the following expression for the transformed metric $\bar{\gamma}^{ij}$
\begin{align}
\label{Conformal_inverse_metric}
\bar{\gamma}^{ij}=\left( \begin{array}{ccc}  \frac{1}{F(\bar{x},\bar{y})} & 0 & 0 \\ 0 & \frac{1}{F(\bar{x},\bar{y})} & 0\\ 0 & 0 & 1 \end{array}\right)
\Rightarrow \sqrt{\bar{\gamma}}\bar{\gamma}^{ij}=\left( \begin{array}{ccc}  1 & 0 & 0 \\ 0 & 1 & 0\\ 0 & 0 & F(\bar{x},\bar{y}) \end{array}\right)
\end{align}
with
\begin{align}
\label{Conformal_factor}
F(\bar{x},\bar{y})=\left(\frac{\partial{x}}{\partial{\bar{x}}}\right)^2+\left(\frac{\partial{y}}{\partial{\bar{x}}}\right)^2
\end{align}
Moreover, we will assume that the pressure $p$ is invariant in the $z$-direction, i.e., $p=p(x,y)$. This implies that
$\gamma^{33}$ is irrelevant, since it is multiplied by $\partial_3p=\partial_z p=0$ and ensure that our problem is effectively isotropic.
As a consequence, taking into account that $\sqrt{\tilde{\gamma}}\tilde{\gamma}^{ij}=\delta^{ij}$, 
equations~\eqref{Wave_equation_pressure_virtual} and \eqref{Wave_equation_pressure_real} are mathematically identical when
\begin{align}
\rho_{\rm R}&=\rho_{\rm V},\\
B_{\rm R}&=\frac{B_{\rm V}}{F(\bar{x},\bar{y})}  \Rightarrow c_{\rm R}^2=\frac{c_{\rm V}^2}{F(\bar{x},\bar{y})}.
\end{align}
This means that we only need to construct a spatially varying bulk modulus to implement a 2D conformal
transformation.

\subsection{2D conformal mappings and static background in ATA}\label{SS:2d_conformal_mappings}
Let us now reanalyze the case of 2D conformal mappings within the ATA framework.
We will assume again that the coordinate system $S_1$ in laboratory space (see Fig.~\ref{fig:Proposal}) is the
Cartesian one ($\gamma_{ij}=\tilde{\gamma}_{ij}=\delta_{ij}$). Thus, we obtain from equation~\eqref{metric_transformation}
that, under the conformal assumption, the transformation given by equation~\eqref{2D_transformation} corresponds to a
metric\footnote{As in the previous section, in our 2D
problem we assume that the velocity potential is invariant in the $z$-direction, so that $\phi=\phi(x,y)$. Taking this fact
into account, from equation~\eqref{dAlembertian} we observe that the value of $g^{33}$ is irrelevant, as it is multiplied by
$\partial_3\phi=\partial_z\phi=0$. Thus, we have an effective 2D isotropic spatial metric, since $g^{11}=g^{22}$.}
\begin{align}
\label{Conformal_direct_metric}
\bar{g}^{\mu\nu}=\frac{1}{\rho_{\rm V}c_{\rm V}}\left( \begin{array}{cccc}  -1 & 0 & 0 & 0 \\ 0 & \frac{c_{\rm V}^2}{F(\bar{x},\bar{y})} & 0 & 0\\ 0 & 0 & \frac{c_{\rm V}^2}{F(\bar{x},\bar{y})} & 0 \\ 0 & 0 & 0 & c_{\rm V}^2\end{array}\right) \Rightarrow \sqrt{-\bar{g}}=\frac{\rho_{\rm V}^2}{c_{\rm V}}F(\bar{x},\bar{y}),
\end{align}
where $F(\bar{x},\bar{y})$ is given by equation~\eqref{Conformal_factor}. On the other hand, using the expression of $\tilde{\gamma}_{ij}$,
\begin{align}
\label{Conformal_physical_metric}
\tilde{g}^{\mu\nu}=\frac{1}{\rho_{\rm R}c_{\rm R}}\left( \begin{array}{cccc}  -1 & 0 & 0 & 0 \\ 0 & c_{\rm R}^2 & 0 & 0\\ 0 & 0 & c_{\rm R}^2 & 0 \\ 0 & 0 & 0 & c_{\rm R}^2\end{array}\right) \Rightarrow \sqrt{-\tilde{g}}=\frac{\rho_{\rm R}^2}{c_{\rm R}}.
\end{align}
Substituting equations~\eqref{Conformal_direct_metric}-\eqref{Conformal_physical_metric} into equation \eqref{ATA_condition},
we obtain the following relations
\begin{align}
\label{Conformal_density}
\rho_{\rm R}&=\rho_{\rm V},\\
\label{Conformal_speed}
c_{\rm R}^2&=\frac{c_{\rm V}^2}{F(\bar{x},\bar{y})}.
\end{align}
As can be seen, for 2D conformal mappings, STA and ATA give the same results.

\subsection{3D conformal mappings and static background in ATA}
We will now consider 3D transformations of a non-moving medium ($\mathbf{v}_{\rm V}=0$). 
In this case,
as in equations~\eqref{metric1} and \eqref{metric2}, we assume (using also equation~\eqref{metric_determinant}) 
\begin{align}
\bar{\gamma}_{ij}&=\Omega_{\rm V}\delta_{ij} \Rightarrow \sqrt{-\bar{g}}=\Omega_{\rm V}^{3/2}\frac{\rho_{\rm V}^2}{c_{\rm V}},  \label{det1} \\
\tilde{\gamma}_{ij}&=\Omega_{\rm R}\delta_{ij} \Rightarrow \sqrt{-\tilde{g}}=\Omega_{\rm R}^{3/2}\frac{\rho_{\rm R}^2}{c_{\rm R}}, \label{det2}
\end{align}
and the metrics in steps 3 and 4 of Fig.~\ref{fig:Proposal} will have the following form

\begin{align}
\label{metric_virtual_conformal}
\bar{g}_{\mu\nu}=\frac{\rho_{\rm V}}{c_{\rm V}}\left( \begin{array}{cccc} -c_{\rm V}^2 & 0 & 0 & 0 \\ 0 & \Omega_{\rm V} & 0& 0\\ 0 & 0 & \Omega_{\rm V} & 0 \\  0 & 0 & 0 & \Omega_{\rm V} \end{array}\right) \Rightarrow
\bar{g}^{\mu\nu}=\frac{1}{\rho_{\rm V}c_{\rm V}}\left( \begin{array}{cccc} -1 & 0 & 0 & 0 \\ 0 & \frac{c_{\rm V}^2}{\Omega_{\rm V}} & 0& 0\\ 0 & 0 & \frac{c_{\rm V}^2}{\Omega_{\rm V}} & 0 \\  0 & 0 & 0 & \frac{c_{\rm V}^2}{\Omega_{\rm V}} \end{array}\right), \\
\label{metric_real_conformal}
\tilde{g}_{\mu\nu}=\frac{\rho_{\rm R}}{c_{\rm R}}\left( \begin{array}{cccc} -c_{\rm R}^2 & 0 & 0 & 0 \\ 0 & \Omega_{\rm R} & 0& 0\\ 0 & 0 & \Omega_{\rm R} & 0 \\  0 & 0 & 0 & \Omega_{\rm R} \end{array}\right) \Rightarrow
\tilde{g}^{\mu\nu}=\frac{1}{\rho_{\rm R}c_{\rm R}}\left( \begin{array}{cccc} -1 & 0 & 0 & 0 \\ 0 & \frac{c_{\rm R}^2}{\Omega_{\rm R}} & 0& 0\\ 0 & 0 & \frac{c_{\rm R}^2}{\Omega_{\rm R}} & 0 \\  0 & 0 & 0 & \frac{c_{\rm R}^2}{\Omega_{\rm R}} \end{array}\right).
\end{align}
Notice that to find $\Omega_{\rm V}$ we must use equation~\eqref{metric_transformation} for a particular transformation.
On the other hand, $\Omega_{\rm R}$ will be determined by the coordinate system $S_1$ employed in laboratory space. 
Using equation~\eqref{ATA_condition} and with the help of equations~\eqref{det1}-\eqref{det2}, we arrive at
\begin{align}
\label{Speed_potential}
c_{\rm R}^2&=c_{\rm V}^2\frac{\Omega_{\rm R}}{\Omega_{\rm V}},\\
\label{Density_potential2}
\rho_{\rm R}&=\rho_{\rm V}\frac{\Omega_{\rm V}^{1/2}}{\Omega_{\rm R}^{1/2}},\\
\label{Bulk_potential2}
B_{\rm R}&=B_{\rm V}\frac{\Omega_{\rm R}^{1/2}}{\Omega_{\rm V}^{1/2}}.
\end{align}

Comparing with equations~\eqref{Bulk_pressure2}-\eqref{Speed_pressure}, it is seen that the transformation of the speed of sound is the same in ATA as the one obtained within STA, whereas the other quantities are transformed differently. In particular, in ATA, one finds that an increase of the speed of sound should correspond to a decrease in the density of the metafluid. As we mentioned above, such properties should be more feasible from a technological point of view than the ones
obtained by STA.

\subsection{General conformal mapping and moving background}
Let us now consider the possibility of a virtual medium with non-zero background velocity in virtual space and try to find out how this velocity should be changed to
implement a certain coordinate transformation. We proceed as in the previous section but considering non-vanishing background velocities. In this
case, the metrics in steps 3 and 4 of Fig. \ref{fig:Proposal}) are
\begin{align}
\bar{g}^{\mu\nu} = \frac{1}{\rho_{\rm V}c_{\rm V}}
\left(
\begin{array}{ccc}
-1 & \vdots & -\bar{v}_{\rm V}^i \\
... & . & ................ \\
-\bar{v}_{\rm V}^i & \vdots & c_{\rm V}^2\Omega_{\rm V}^{-1}\delta^{ij}-\bar{v}_{\rm V}^i\bar{v}_{\rm V}^j
\end{array}
\right), \\
\tilde{g}^{\mu\nu} = \frac{1}{\rho_{\rm R}c_{\rm R}}
\left(
\begin{array}{ccc}
-1 & \vdots & -\tilde{v}_{\rm R}^i \\
... & . & ................ \\
-\tilde{v}_{\rm R}^i & \vdots & c_{\rm R}^2\Omega_{\rm R}^{-1}\delta^{ij}-\tilde{v}_{\rm R}^i\tilde{v}_{\rm R}^j
\end{array}
\right).
\end{align}
Clearly, the only additional requirement to satisfy equation~\eqref{ATA_condition} (along with
equations~\eqref{Speed_potential}-\eqref{Density_potential2}) is that $\tilde{v}_{\rm R}^i=\bar{v}_{\rm V}^i$,
and we know from equation~\eqref{metric_transformation} that for pure spatial transformations
\begin{align}\label{back_velocity}
\bar{v}_{\rm V}^{\bar{i}}=\bar{g}^{\bar{0}\bar{i}}=\Lambda^{\bar{0}}_{0}\Lambda^{\bar{i}}_{i}g^{0i}=\Lambda^{\bar{i}}_{i}v_{\rm V}^{i}= \frac{\partial{\bar{x}^{\bar{i}}}}{\partial{x^i}}v_{\rm V}^{i}.
\end{align}
Therefore, differently from STA, which assumes a zero background velocity, ATA allows us to implement transformations of moving media and provides us with the background velocity field required to achieve a certain transformation. This is one of the main advantages of ATA.

\section{Spacetime transformations}
At this point we are ready to consider transformations that mix space and time in ATA, which could not be implemented
in STA. We will start from a virtual medium $M_1$ characterized by parameters $\rho_{\rm V}$, $c_{\rm V}$
and $\mathbf{v}_{\rm V} = 0$ (Step 1 in Fig. \ref{fig:Proposal}), and use a Cartesian system of coordinates (system $S_1$)
in laboratory space. According to equation~\eqref{inverse_4metric}, the equivalent metric of the equation in step 2
of Fig. \ref{fig:Proposal} for this configuration is
\begin{align}
g^{\mu\nu}=\frac{1}{\rho_{\rm V}c_{\rm V}}\left( \begin{array}{cccc} -1 & 0 & 0 & 0 \\ 0 & c_{\rm V}^2 & 0& 0\\ 0 & 0 & c_{\rm V}^2 & 0 \\  0 & 0 & 0 & c_{\rm V}^2 \end{array}\right)
\Rightarrow \sqrt{-g}=\frac{\rho_{\rm V}^2}{c_{\rm V}}.
\end{align}
On the other hand, we consider a real medium (Step 5 in Fig. \ref{fig:Proposal}), moving with background 
velocity $\mathbf{v}_{\rm R}$ and characterized by isotropic density and sound speed $\rho_{\rm R}$ and $c_{\rm R}$. We also use the Cartesian
system $S_1$ in this case, and thus the metric associated to Equation 4 in Fig. \ref{fig:Proposal} is given by

\begin{equation}
\tilde{g}^{\mu\nu}(t,\x) \equiv 
{1\over \rho_{\rm R} c_{\rm R}}
\left( \begin{matrix}-1&\vdots&-v_{\rm R}^j\cr
               \cdots\cdots&\cdot&\cdots\cdots\cdots\cdots\cr
	       -v_{\rm R}^i&\vdots&(c_{\rm R}^2 \; \delta^{ij} - v_{\rm R}^i \; v_{\rm R}^j )\cr \end{matrix} 
\right)
\Rightarrow \sqrt{-\tilde{g}}=\frac{\rho_{\rm R}^2}{c_{\rm R}}.
\end{equation}

By way of example, next we will analyze two different cases.

\subsection{Example 1: three-dimensional dynamic contraction}\label{SS:dynamic_contraction}
Consider the transformation
\begin{align}
\bar{t}&=t,\\
\bar{x}&=xf_0(t),\\
\bar{y}&=yf_0(t),\\
\bar{z}&=zf_0(t).
\end{align}
The transformation matrix in this case is
\begin{align}
\Lambda^{\bar{\mu}}_{\mu}=\left( \begin{array}{cccc} 1 & 0 & 0 & 0 \\ xf_0'(t) & f & 0& 0\\ yf_0'(t) & 0 & f & 0 \\  zf_0'(t) & 0 & 0 & f \end{array}\right).
\end{align}
Thus, the transformed metric is
\begin{equation}
\bar{g}^{\mu\nu}(t,\x) \equiv 
{1\over \rho_{\rm V} c_{\rm V}}
\left( \begin{matrix}-1&\vdots&-x_jf_0'(t)\cr
               \cdots\cdots&\cdot&\cdots\cdots\cdots\cdots\cr
	       -x_if_0'(t)&\vdots&c_{\rm V}^2f_0^2(t) \; \delta^{ij} - x_i \; x_j \; f_0'^2(t) \cr \end{matrix} 
\right)
\Rightarrow \sqrt{-\bar{g}}=\frac{\rho_{\rm V}^2}{c_{\rm V}f_0^3(t)},\nonumber
\end{equation}
where the prime represents the derivative with respect to time. 
Imposing equation~\eqref{ATA_condition}, the following real medium parameters are obtained
\begin{align}
v_{\rm R}^i&=x^if_0'(t),\\
c_{\rm R}&=c_{\rm V}f_0(t),\\
\rho_{\rm R}&=\frac{\rho_{\rm V}}{f_0(t)}.
\end{align}

\subsection{Example 2: General transformation mixing time and one spatial dimension}\label{SS:general_mixing}
We consider now the most general transformation mixing one spatial coordinate and time:
\begin{align}
\bar{t}&=f_1(x,t),\\
\bar{x}&=f_2(x,t),\\
\bar{y}&=y,\\
\bar{z}&=z.
\end{align}
The transformation matrix in this case is
\begin{align}
\Lambda^{\bar{\mu}}_{\mu}=\left( \begin{array}{cccc} \partial_t{f}_1 & \partial_x{f}_1 & 0 & 0 \\ \partial_t{f}_2 & \partial_x{f}_2 & 0& 0\\ 0 & 0 & 1 & 0 \\  0 & 0 & 0 & 1 \end{array}\right).
\end{align}
Therefore, the transformed metric is
\begin{align}
\bar{g}^{\mu\nu}&=\frac{1}{\rho_{\rm V}c_{\rm V}}\left( \begin{array}{cccc} -(\partial_t{f}_1)^2+c_{\rm V}^2(\partial_x{f}_1)^2 & -\partial_t{f}_1\partial_t{f}_2+c_{\rm V}^2\partial_x{f}_1\partial_x{f}_2 & 0 & 0 \\ -\partial_t{f}_1\partial_t{f}_2+c_{\rm V}^2\partial_x{f}_1\partial_x{f}_2 & -(\partial_t{f}_2)^2+c_{\rm V}^2(\partial_x{f}_2)^2 & 0& 0\\ 0 & 0 & c_{\rm V}^2 & 0 \\  0 & 0 & 0 & c_{\rm V}^2 \end{array}\right), \\
\sqrt{-\bar{g}}&=\frac{\rho_{\rm V}^2}{c_{\rm V}}\frac{1}{\partial_t{f}_1\partial_x{f}_2-\partial_x{f}_1\partial_t{f}_2}.
\end{align}
Imposing equation~\eqref{ATA_condition}, we obtain the parameters of the real medium that reproduce this transformation
\begin{align}
v_{\rm R}^x&=\frac{\partial_t{f}_1\partial_t{f}_2-c_{\rm V}^2\partial_x{f}_1\partial_x{f}_2}  {(\partial_t{f}_1)^2-c_{\rm V}^2(\partial_x{f}_1)^2},\\
c_{\rm R}^2&=(v_{\rm R}^x)^2 + \frac{c_{\rm V}^2(\partial_x{f}_2)^2-(\partial_t{f}_2)^2} {(\partial_t{f}_1)^2-c_{\rm V}^2(\partial_x{f}_1)^2},\\
\rho_{\rm R}&=\rho_{\rm V}\frac{c_{\rm R}^2}{c_{\rm V}^2} \frac {(\partial_t{f}_1)^2-c_{\rm V}^2(\partial_x{f}_1)^2} {\partial_t{f}_1\partial_x{f}_2-\partial_x{f}_1\partial_t{f}_2}.
\end{align}

We stress again that the transformations presented in this section were not possible in STA.

\subsection{Transformations for the spacetime cloak}
{Here we include the transformation used to design the spacetime cloak described in the main text.
The transformation consists of the composition of a Lorentz boost
\begin{align}
x_1&=(1-n^{-2})^{-1/2}\left(x-\frac{ct}{n}\right)\\
t_1&=(1-n^{-2})^{-1/2}\left(t-\frac{x}{nc}\right)
\end{align}
followed by a curtain map
\begin{align}
x_2&=\left(\frac{\delta+|ct_1|}{\delta+n\sigma}\right)\left[x_1-{\rm sgn}(x_1)\sigma\right]+{\rm sgn}(x_1)\sigma\\
t_2&=t_1
\end{align}
and an inverse Lorentz boost
\begin{align}
\bar{x}&=(1-n^{-2})^{-1/2}\left(x_2+\frac{ct_2}{n}\right)\\
\bar{t}&=(1-n^{-2})^{-1/2}\left(t_2+\frac{x_2}{nc}\right)
\end{align}
In our example $\sigma=1$, $n=2$ and $\delta=0.5$.}
\section{Numerical examples}

In this section we present some numerical examples for each of the cases studied in the previous sections.

\subsection{2D conformal transformation}

\begin{figure}[!ht] 
\begin{center}
  \includegraphics{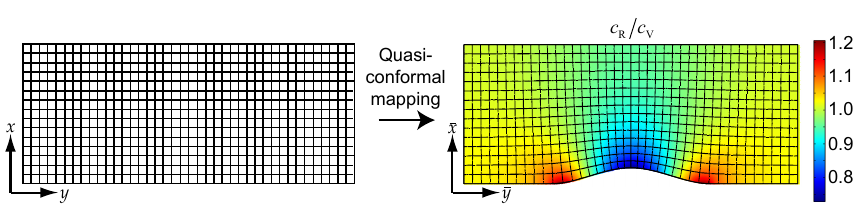}
\end{center}
  \caption{Carpet cloak quasi-conformal mapping.}
  \label{fig:CarpetCloakMapping}
\end{figure}

As a first verification of the ATA approach based on the velocity potential equation, we designed and simulated a so-called carpet
cloak, a concept that originally appeared in the framework of transformation optics~\cite{LI08-PRL}. Essentially, a carpet cloak makes a curved
wall look flat for an external observer. Thus, we can create a bump in the wall and hide an object behind it without producing
additional scattering. The advantage of the carpet cloak is that it can be obtained from a quasi-conformal transformation
that introduces a very small amount of anisotropy, which can essentially be neglected. To our knowledge, acoustic carpet cloaks based on quasi-conformal mappings have not been reported so far. On the other hand, an anisotropic version of this acoustic device was recently demonstrated
both theoretically and experimentally~\cite{POP11-PRL}. Here, we design and numerically verify an isotropic version of the
acoustic carpet cloak. To obtain the mapping, we will use the method described in Ref.~\cite{CHANG}, with which we can calculate
the function F($\bar{x}$,$\bar{y}$). The idea is to deform a rectangular region as shown in Fig.~\ref{fig:CarpetCloakMapping}.
Using the above-mentioned algorithm and equation~\eqref{Conformal_speed}, we calculated the speed of sound distribution associated
to this transformation, which is also depicted in Fig.~\ref{fig:CarpetCloakMapping}.

\begin{figure}
\begin{center}
  \includegraphics{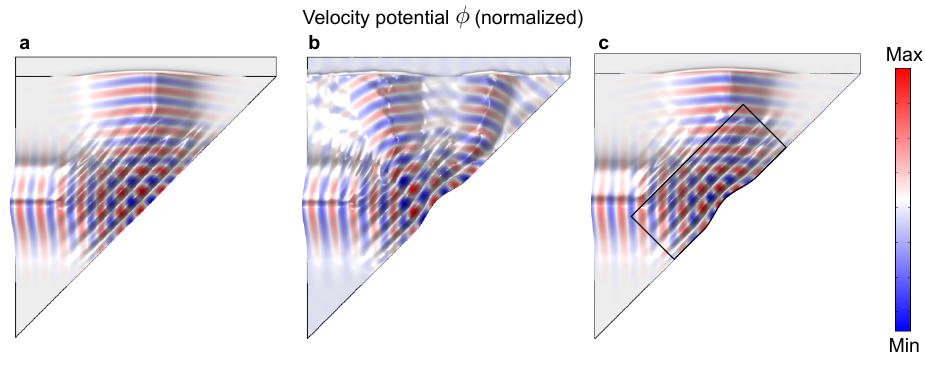}
\end{center}
  \caption{Simulated velocity potential (color) of a Gaussian beam impinging from the left onto different objects. (a) Flat wall. (b) Wall with a bump.
(c) Wall with a bump surrounded by a carpet cloak. The figure height for which lighting is calculated corresponds to wave intensity.}
  \label{fig:CarpetCloakIntensity}
\end{figure}

To test the device, we simulated its performance both starting from the velocity potential wave equation and 
from the pressure wave equation with the same sound speed distribution, obtaining identical results,
as expected based on the conclusions of Section~\ref{SS:2d_conformal_mappings}.
The results for the first case are shown in Fig.~\ref{fig:CarpetCloakIntensity}, where we compare the
velocity potential associated to a Gaussian beam impinging from the left onto a flat wall, a wall with a bump,
and a wall with a bump surrounded by a carpet cloak.

\subsection{3D conformal transformation and Maxwell's fish-eye}

\begin{figure}
\begin{center}
  \includegraphics{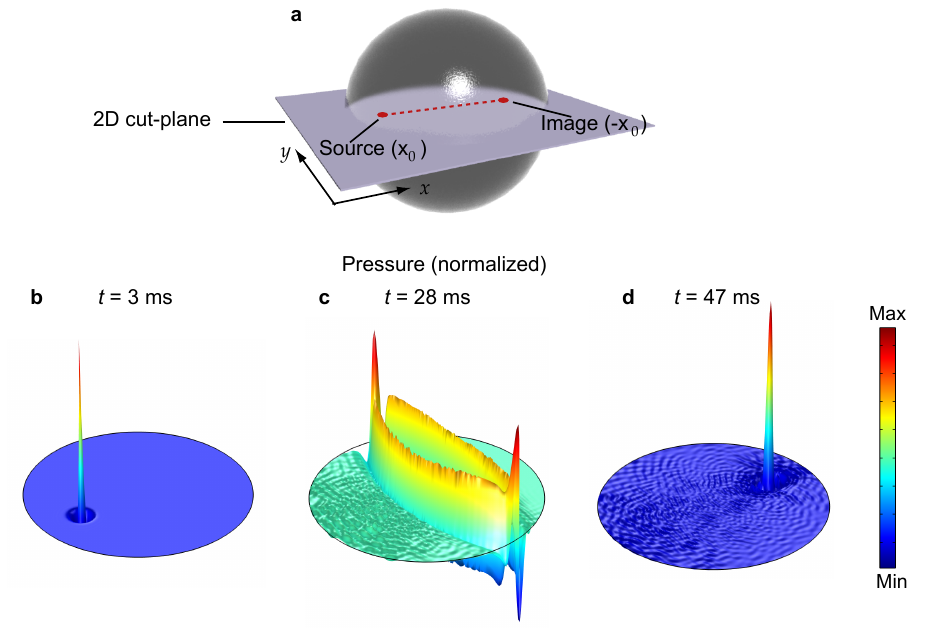}
\end{center}
  \caption{3D simulation of Maxwell's fish-eye under point source excitation. (a) Device schematic. In the simulation, $a=5$ m and
$\mathbf{x}_0=-3\hat{x}$ m. The structure is surrounded by a hard boundary condition (wall) that acts as a mirror.
A point source emits a Gaussian pulse (in the time domain) at a certain point within the spherical structure ($\mathbf{x}_0$).
We keep track of this wave at different instants to verify the focusing effect. In the figure, we depict the acoustic
pressure over a spherical cross-section passing through its center at (b) $t$ = 3 ms, when the pulse has just been launched,
(c) $t$ = 28 ms, and (d) $t$ = 47 ms, when the pulse gets focused at $-\mathbf{x}_0$. Note that, although we only show a 2D plane here,
 the focusing effect occurs in 3D.}
  \label{fig:Fisheye3D}
\end{figure}

\begin{figure}
\begin{center}
  \includegraphics{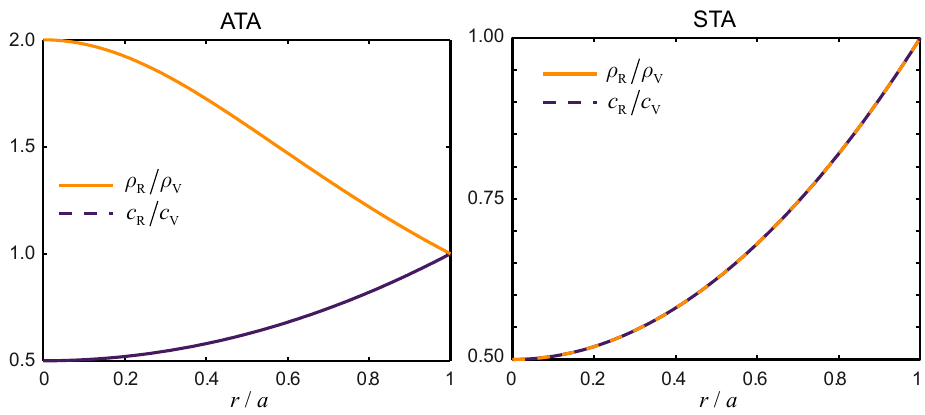}
\end{center}
  \caption{3D Maxwell's fish-eye mass density and sound speed as a function of the radius.
As can be seen, these two parameters follow an inverse behavior in the analogue approach (left), whereas they had the same behavior in the standard approach (right).}
  \label{fig:Fisheye_properties}
\end{figure}

A very interesting example of an isotropic 3D device is that of Maxwell's fish-eye. This element implements the
mapping of the 3D surface of the four-dimensional hypersphere to the volume of a 3D sphere in flat 3D space~\cite{LEO,LEO10-PRB}.
The metric of such a geometry in Cartesian coordinates is given by~\cite{LEO}
\begin{align}
\label{Fisheye_metric}
\left(\bar{g}_{\mu\nu}\right)=\frac{\rho_{\rm V}}{c_{\rm V}}\left( \begin{array}{cccc} -c_{\rm V}^2 & 0 & 0 & 0 \\ 0 & \Omega_{\rm V} & 0& 0\\ 0 & 0 & \Omega_{\rm V} & 0 \\  0 & 0 & 0 & \Omega_{\rm V} \end{array}\right),
\end{align}
with
\begin{align}
\Omega_{\rm V}=\left(\frac{2}{1+\frac{r^2}{a^2}}\right)^2, \qquad r=\sqrt{x^2+y^2+z^2},
\end{align}
where $a$ is the radius of the 4-sphere. Therefore, the device has an infinite extension in principle. However, it can be shown
that its functionality is unaltered if we place a spherical mirror at $r=a$. One of the main properties of Maxwell's fish-eye is that it focuses
the rays emanating from a point source located at any arbitrary point $\mathbf{x}_0$ within the sphere to the point $-\mathbf{x}_0$~\cite{LEO,LEO10-PRB}.
Of course, this is not valid for points outside the spherical mirror.
According to equations~\eqref{Speed_potential}-\eqref{Density_potential2}, we can implement the metric given by equation~\eqref{Fisheye_metric}
by using a medium with the following properties ($\Omega_{\rm R}=1$)

\begin{align}
\label{Bulk_potential_fisheye}
c_{\rm R}&=c_{\rm V}\left(\frac{2}{1+\frac{r^2}{a^2}}\right)^{-1},\\
\label{Density_potential_fisheye}
\rho_{\rm R}&=\rho_{\rm V}\frac{2}{1+\frac{r^2}{a^2}}.
\end{align}
We have verified the functionality of the acoustic Maxwell's fish-eye through 3D numerical calculations as shown in
Fig.~\ref{fig:Fisheye3D}, where we can observe its focusing properties.
Being a 3D device, the properties required to implement Maxwell's fish-eye derived from the ATA approach could be
easier to achieve than those derived from the STA approach, as has been discussed above (see Fig.~\ref{fig:Fisheye_properties}).

\subsection{Transforming the background velocity}

\begin{figure}
\begin{center}
  \includegraphics{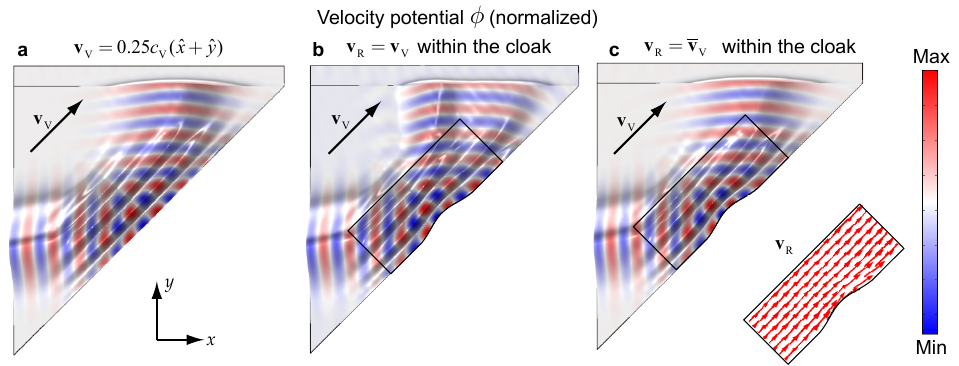}
\end{center}
  \caption{(a) Flat wall with a constant background velocity parallel to the wall. (b) Carpet cloak with the same background velocity as in (a).
(c) Carpet cloak with a properly transformed background velocity. The figure height for which lighting is calculated corresponds to wave intensity.}
  \label{fig:AeroacousticCarpetCloak}
\end{figure}

To verify the possibility of working with moving media provided by our new approach, we will use the previous
example of the carpet cloak, but considering a non-zero background velocity parallel to the wall. For instance,
let us take $\mathbf{v}_{\rm V}=0.25c_{\rm V}(\hat{x}+\hat{y})$ m/s. If we use the values of $\rho_{\rm R}$ and $c_{\rm R}$
calculated for the carpet cloak in the previous section and keep this value of $\mathbf{v}_{\rm V}$, the wave is
distorted into an undesired wave (see Fig.~\ref{fig:AeroacousticCarpetCloak}). However, if we force a background
velocity given by equation~\eqref{back_velocity} such that
\begin{align}
\left( \begin{array}{c} \tilde{v}^x_{\rm R} \\ \tilde{v}^y_{\rm R} \end{array}\right)
=\left( \begin{array}{cc} \frac{\partial{\bar{x}}}{\partial{x}} & \frac{\partial{\bar{x}}}{\partial{y}} \\ \frac{\partial{\bar{y}}}{\partial{x}} & \frac{\partial{\bar{y}}}{\partial{y}} \end{array}\right)
\left( \begin{array}{c} v^x_{v} \\ v^y_{v} \end{array}\right),
\end{align}
the wave is distorted in the desired way (remains unchanged outside the cloak).

\subsection{3D dynamic contraction}
In this example we simulate the transformation of section~\ref{SS:dynamic_contraction} in the ray acoustics approximation
(for which only the sound speed and background velocity are relevant).
The transformation is performed only inside a 3D box, which we call the time-dependent spatial compressor. 
The employed function $f_0(t)$ is shown in Fig. 2 of the main text. When $f_0(t) \neq 1$, 
the transformation is discontinuous at the boundaries of the compressor, so the ray must enter and exit the compressor
when $f_0(t)=1$ in order to avoid reflections.

\begin{figure}
\begin{center}
  \includegraphics{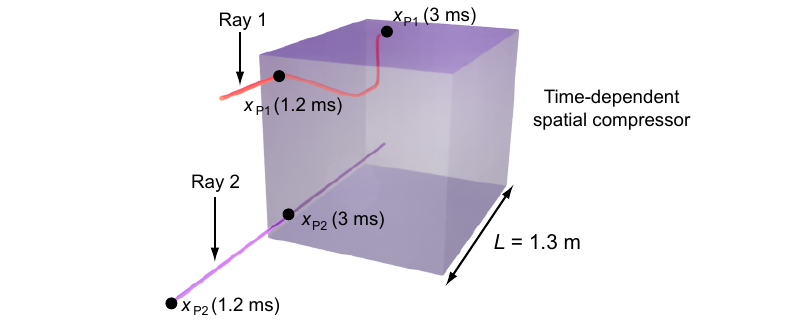}
\end{center}
  \caption{Time-dependent spatial compressor. The box has a length of $L=1.3$ m. The positions $\mathbf{x}_{\rm P1}$ and
$\mathbf{x}_{\rm P2}$ of the first and second acoustic rays at different instants are shown. When the compression starts ($t=1.2$ ms),
Ray 1 has just entered the compressor. When the compression finishes ($t=3$ ms), Ray 1 is about to exit
the box and Ray 2 is about to enter.}
  \label{fig:Spacetime_compressor}
\end{figure}

In Fig.~\ref{fig:Spacetime_compressor} we depict the simulated trajectories followed by two different rays.
After Ray 1 has just entered the box, the compression starts. It reaches a value of $f_0(t)=0.5$, which is kept during a
short time interval, and finally returns to no compression ($f_0(t)=1$) before Ray 1 exits the box. Ray 2
feels no compression since $f_0(t)=1$ during the time it goes through the box. Thus, Ray 2 is a straight line.

We can use this compressor, for instance, to select which rays are trapped by an omnidirectional absorber placed inside the box
(see Fig. 2 in the main text). We take this absorber to be the one described in~\cite{CLI12-APL}, which is characterized by an acoustic
refractive index
\begin{align}
n=\frac{R}{r}n_{\rm s},
\end{align}
where $R$ is the radius of the absorber and $n_{\rm s}$ is the refractive index of the medium surrounding the absorber. In our particular
implementation, we take $R=0.29$ m and $n_{\rm s}=0.5n_{\rm b}$, which is the effective refractive index of the compressor for maximum 
compression ($n_{\rm b}$ is the index of the background fluid).

\section*{Acknowledgements}
This work was developed under the framework of the ARIADNA contract 4000104572/11/NL/KML of the European Space
Agency. A.~M. and J.~S.-D. also acknowledge support from Consolider EMET project (CSD2008-00066), A.~M. from project
TEC2011-28664-C02-02, and C.~B. and G.~J. from the project FIS2008-06078-C03-01. 
We thank Reme Miralles for her help with Fig.~\ref{fig:Figure_2}.

\section*{Author Contributions}
S. C. proposed to explore the combination of transformation acoustics and analogue gravity.
C. G.-M. devised the final method, developed the main theory, and performed the simulations.
C. B., G. J., and S. C. contributed in several important theoretical aspects of this work.
C. G.-M. and J. S.-D. analyzed the implementation advantages of the proposed method.
A. M. assisted with the simulations.
All authors participated in the manuscript preparation and discussed the results.
S. C. and A. M. coordinated the work.


\begin{thebibliography}{99}
\bibitem{PEN06-SCI}
Pendry, J. B., Schurig, D. \& Smith, D. R. Controlling electromagnetic fields. \emph{Science} {\bf 312}, 1780--1782 (2006).
\bibitem{LEO06-SCI}
Leonhardt, U. Optical conformal mapping. \emph{Science} {\bf 312}, 1777--1780 (2006).
\bibitem{SCH06-SCI}
Schurig, D. \emph{et al.} Metamaterial Electromagnetic Cloak at Microwave Frequencies. \emph{Science} {\bf 314}, 977--980 (2006).
\bibitem{SHA08-SCI}
Shalaev, V. M. Transforming Light. \emph{Science} {\bf 322}, 384--386 (2008).
\bibitem{GRE09-BAMS}
Greenleaf, A., Kurylev, Y., Lassas, M. \& Uhlmann, G. Invisibility and inverse problems. \emph{B. Am. Math. Soc.} {\bf 46}, 55--97 (2009).
\bibitem{GEN09-NP}
Genov, D. A., Zhang, S. \& Zhang, X. Mimicking celestial mechanics in metamaterials. \emph{Nat. Phys.} {\bf 5}, 687--692 (2009).
\bibitem{CHE10-NM}
Chen, H., Chan, C. T. \& Sheng, P. Transformation optics and metamaterials. \emph{Nat. Mater.} {\bf 9}, 387--396 (2010).
\bibitem{LEO}
Leonhardt, U. \& Philbin, T. \emph{Geometry and light. The science of invisibility} (Dover Publications, 2010).
\bibitem{PEN12-SCI}
Pendry, J. B., Aubry, A., Smith, D. R. \& Maier, S. A. Transformation Optics and Subwavelength Control of Light. \emph{Science} {\bf 337}, 549 (2012).
\bibitem{POST}
Post, E. G. \emph{Formal Structure of Electromagnetics: General Covariance and Electromagnetics} (Interscience Publishers, New York, 1962).
\bibitem{CUM07-NJP}
Cummer, S. A. \& Schurig, D. One path to acoustic cloaking. \emph{New J. Phys.} {\bf 9}, 45 (2007).
\bibitem{CHE07-APL}
Chen, H. \& Chan, C. T. Acoustic cloaking in three dimensions using acoustic metamaterials. \emph{Appl. Phys. Lett.} {\bf 91}, 183518 (2007).
\bibitem{NOR08-JASA}
Norris, A. N. Acoustic metafluids. \emph{J. Acoust. Soc. Am.} {\bf 125}, 839 (2009).
\bibitem{CHE10-JPDAP}
Chen, H. and Chan, C. T. Acoustic cloaking and transformation acoustics. \emph{J. Phys. D: Appl. Phys.} {\bf 43}, 113001 (2010).
\bibitem{ZHA08-PRL}
Zhang, S., Genov, D. A., Sun, C. \& Zhang, X. Cloaking of Matter Waves. \emph{Phys. Rev. Lett.} {\bf 100}, 123002 (2008).
\bibitem{MCC11-JOPT}
McCall, M. W., Favaro, A., Kinsler, P. \& Boardman, A. A spacetime cloak, or a history editor. \emph{J. Opt.} {\bf 13}, 024003 (2011).
\bibitem{FRI12-NAT}
Fridman, M., Farsi, A., Okawachi, Y. \& Gaeta, A. L. Demonstration of temporal cloaking. \emph{Nature} {\bf 481}, 62–65 (2012).
\bibitem{CUM11-JOPT}
Cummer, S. A. \& Thompson, R. T. Frequency conversion by exploiting time in transformation optics. \emph{J. Opt.} {\bf 13}, 024007 (2011).
\bibitem{BAR}
Barcel\'{o}, C., Liberati, S. \& Visser, M. Analogue Gravity. \emph{Living Rev. Relativity} {\bf 14}, 3 (2011).
\bibitem{BER46-JASA}
Bergmann, P. G. The Wave Equation in a Medium with a Variable Index of Refraction. \emph{J. Acoust. Soc. Am.} {\bf 17}, 329 (1946).
\bibitem{TOR06-PRL}
Torrent, D., H\aa kansson, A., Cervera, F. \& S\'{a}nchez-Dehesa, J. Homogenization of two-dimensional clusters of rigid rods in air. \emph{Phys. Rev. Lett.} {\bf 96}, 204302 (2006).
\bibitem{TOR06-PRB}
Torrent, D. \& S\'{a}nchez-Dehesa, J. Effective parameters of clusters of cylinders embedded in a nonviscous fluid or gas. \emph{Phys. Rev. B} {\bf 74}, 224305 (2006).
\bibitem{Unruh}
Unruh, W. G. Experimental black hole evaporation? \emph{Phys.\ Rev.\ Lett.}  {\bf 46}, 1351 (1981).
\bibitem{Visser} 
Visser, M. Acoustic black holes: Horizons, ergospheres, and Hawking radiation. \emph{Class.\ Quant.\ Grav.} {\bf 15}, 1767 (1998).
\bibitem{LI08-PRL}
Li, J. \& Pendry, J. B. Hiding under the Carpet: A New Strategy for Cloaking. \emph{Phys. Rev. Lett.} {\bf 101}, 203901 (2008).
\bibitem{POP11-PRL}
Popa, B. I., Zigoneanu, L. \& Cummer, S. A. Experimental acoustic ground cloak in air. \emph{Phys. Rev. Lett.} {\bf 106}, 253901 (2011).
\bibitem{garay}
Garay, L. J., Anglin, J. R., Cirac J. I. \& Zoller, P.
Black holes in Bose-Einstein condensates. \emph{Phys.\ Rev.\ Lett.} {\bf 85}, 4643 (2000).
\bibitem{steinhauer}
Lahav, O., Itah, A., Blumkin, A., Gordon, C. \& Steinhauer, J.
Realization of a sonic black hole analogue in a Bose-Einstein condensate. \emph{Phys.\ Rev.\ Lett.} {\bf 105}, 240401 (2010).
\bibitem{castro} 
Castro Neto, A. H., Guinea, F., Peres, N. M. R., Novoselov, K. S. \& Geim, A. K. The electronic properties of graphene. \emph{Rev.\ Mod.\ Phys.}  {\bf 81}, 109 (2009).
\bibitem{cortijo}
Cortijo, A. \& Vozmediano, M. A. H. Electronic properties of curved graphene sheets. \emph{Europhys.\ Lett.} {\bf 77}, 47002 (2007).
\bibitem{VAK11-SCI}
Vakil, A. \& Engheta, N. Transformation optics using graphene. \emph{Science} {\bf 332}, 1291 (2011).
\bibitem{Unruh}
Unruh, W. G. Experimental black hole evaporation? \emph{Phys.\ Rev.\ Lett.}  {\bf 46}, 1351 (1981).
\bibitem{Visser} 
Visser, M. Acoustic black holes: Horizons, ergospheres, and Hawking radiation. \emph{Class.\ Quant.\ Grav.} {\bf 15}, 1767 (1998).
\bibitem{BAR}
Barcel\'{o}, C., Liberati, S. \& Visser, M. Analogue Gravity. \emph{Living Rev. Relativity} {\bf 14}, 3 (2011).
\bibitem{TOR09-PRB}
Torrent, D. \& S\'{a}nchez-Dehesa, J. Sound scattering by anisotropic metafluids based on two-dimensional sonic crystals. \emph{Phys. Rev. B} {\bf 79}, 174104 (2009).
\bibitem{TOR06-PRL}
Torrent, D., H\aa kansson, A., Cervera, F. \& S\'{a}nchez-Dehesa, J. Homogenization of two-dimensional clusters of rigid rods in air. \emph{Phys. Rev. Lett.} {\bf 96}, 204302 (2006).
\bibitem{TOR06-PRB}
Torrent, D. \& S\'{a}nchez-Dehesa, J. Effective parameters of clusters of cylinders embedded in a nonviscous fluid or gas. \emph{Phys. Rev. B} {\bf 74}, 224305 (2006).
\bibitem{LI08-PRL}
Li, J. \& Pendry, J. B. Hiding under the Carpet: A New Strategy for Cloaking. \emph{Phys. Rev. Lett.} {\bf 101}, 203901 (2008).
\bibitem{POP11-PRL}
Popa, B. I., Zigoneanu, L. \& Cummer, S. A. Experimental acoustic ground cloak in air. \emph{Phys. Rev. Lett.} {\bf 106}, 253901 (2011).
\bibitem{CHANG}
Chang, Z., Zhou, X., Hu, J. \& Hu, G. Design method for quasi-isotropic transformation materials based on inverse Laplace’s equation with sliding boundaries. \emph{Opt. Express} {\bf 18}, 6089-6096 (2010).
\bibitem{LEO}
Leonhardt, U. \& Philbin, T. \emph{Geometry and light. The science of invisibility} (Dover Publications, 2010).
\bibitem{LEO10-PRB}
Leonhardt, U. \& Philbin, T. G. Perfect imaging with positive refraction in three dimensions, \emph{Phys. Rev. A} {\bf 81}, 011804(R) (2010).
\bibitem{CLI12-APL}
Climente, A., Torrent, D. \& S\'{a}nchez-Dehesa, J. Omnidirectional broadband acoustic absorber based on metamaterials. \emph{Appl. Phys. Lett.} {\bf 100}, 144103 (2012).
\end{thebibliography}
\end{document}